\documentclass[aps,prl,twocolumn,preprintnumbers,groupedaddress,superscriptaddress,floatfix,tightenlines,reprint,nofootinbib]{revtex4-1}
\usepackage{mathrsfs,natbib}
\usepackage{graphics,epsfig,color,graphicx}
\usepackage{verbatim}
\usepackage{bm}
\usepackage{graphicx,epsf,amssymb,amsbsy,amsfonts,amssymb,amsmath}
\usepackage{hyperref}

\newcommand{\be}{\begin{eqnarray}}
\newcommand{\ee}{\end{eqnarray}}
\newcommand{\beq}{\begin{equation}}
\newcommand{\eeq}{\end{equation}}

\newcommand{\nl}{\nonumber \\}

\newcommand{\Tr}{{\rm Tr}}
\newcommand{\tr}{{\rm tr}}

\usepackage[T1]{fontenc} % if needed
\usepackage{slashed}
\usepackage{mathtools}

\def\tr{\text{tr}\,}

\hypersetup{
    colorlinks=true,    % false: boxed links; true: colored links
    linkcolor=red,      % color of internal links 
    citecolor=blue,     % color of links to bibliography
    filecolor=magenta,  % color of file links
    urlcolor=blue       % color of external links
}

\makeatletter
\newcommand{\biggg}{\bBigg@\thr@@}
\newcommand{\Biggg}{\bBigg@{3.0}}

\makeatother

\begin{document}
%%%%%%%%%%%%%%%%%%%%%%%%%%%%%%%%%%%%%%%%%%%%%%%%%%%%%%%%%%%%%%%%%%%%%%%%%%%

\title{Flow of Hagedorn singularities and phase transitions in large $N$
  gauge theories}
 
\author{Aleksey Cherman}
\email{acherman@umn.edu}
\affiliation{School of Physics and Astronomy, University of Minnesota
  Minneapolis, MN 55455}

\author{Syo Kamata}
\email{skamata11phys@gmail.com}
\affiliation{College of Physics and Communication Electronics,
  Jiangxi Normal University,\\
  Nanchang 330022, China}
\affiliation{Department of Physics, North Carolina State University,
  Raleigh, NC 27695, USA}

\author{Thomas Sch\"afer}
\email{tmschaef@ncsu.edu}
\affiliation{Department of Physics, North Carolina State University,
  Raleigh, NC 27695, USA}

\author{Mithat \"Unsal}
\email{unsal.mithat@gmail.com}
\affiliation{Department of Physics, North Carolina State University,
  Raleigh, NC 27695, USA}

\begin{abstract}
  We investigate the singularity structure of the $(-1)^F$ graded partition 
function in QCD with $n_f \geq 1$ massive adjoint fermions in the
large-$N$ limit. Here, $F$ is fermion number and $N$ is the number of
colors. The large $N$ partition function is made reliably calculable
by taking space to be a small three-sphere $S^3$. Singularites in the
graded partition function are related to phase transitions and to Hagedorn
behavior in the $(-1)^F$-graded density of states.  We study the flow of
the singularities in the complex ``inverse temperature'' $\beta$  plane
as a function of the quark mass. This analysis is a generalization of
the Lee-Yang-Fisher-type analysis for a theory which is always in the
thermodynamic limit thanks to the large $N$ limit. We identify two
distinct mechanisms for the appearance of physical Hagedorn singularities
and center-symmetry changing phase transitions at real positive $\beta$,
inflow of singularities from the $\beta=0$ point, and collisions of
complex conjugate pairs of singularities. 
\end{abstract}

\maketitle

%%%%%%%%%%%%%%%%%%%%%%%%%%%%%%%%%%%%%%%%%%%%%%%%%%%%%%%%%%%%%%%%%%%%%%%%%%
{\bf Introduction:}
%%%%%%%%%%%%%%%%%%%%%%%%%%%%%%%%%%%%%%%%%%%%%%%%%%%%%%%%%%%%%%%%%%%%%%%%%%
Singularities of partition functions play an important role in the
study of phase transitions. Important examples that have been discussed
in the literature are Lee-Yang zeros, Fisher zeros and Hagedorn
singularities. Lee and Yang studied zeros of the partition function
of a finite system in the presence of an external field \cite{Lee:1952}.
The nature of phase transitions is controlled by the motion of 
singularities as the system approaches the thermodynamic limit,
along with the limiting distribution of the Lee-Yang zeros. Fisher studied
zeros of the partition function as a function of a complex temperature or
interaction parameter \cite{Fisher:1965}.  Hagedorn singularities are
generally a signature of confinement in large $N$ gauge theories, see
e.g.~\cite{Hagedorn:1965st,Cohen:2009wq,Cohen:2011yx}.  They are produced
by exponential growth in the number of states as a function of energy, which
is believed to be related to the excitation spectrum of confining strings.
The existence of these singularities in the thermal partition function is
tied to the inevitability of a deconfinement  phase transition as a function
of temperature in e.g. large $N$ Yang-Mills theory \cite{Cabibbo:1975ig}.

 In this work we investigate the motion of partition function
singularities of large $N$ gauge theories on compactified spaces. There 
are several novel aspects compared to the works of Lee, Yang and Fisher. 
First, gauge theories with a large number of colors $N$ can have phase
transitions even in a finite volume, because the large $N$ limit is
basically a thermodynamic limit \cite{Yaffe:1981vf}. In the present work
we take advantage of this fact by using a finite spatial volume to study
phase transitions at weak gauge coupling. We work in Euclidean spacetime
signature, take space to be a three-sphere $S^3$ with radius $R$, and assume
that Euclidean time is a circle $S^1$ of circumference $\beta$. For an
asymptotically free theory with strong scale $\Lambda$, the limit
$R\Lambda \ll 1$ is a weak-coupling limit.  Gauge theories in this limit
were first studied in
Refs.~\cite{Sundborg:1999ue,Polyakov:2001af,Aharony:2003sx}. 

Our focus will be on the singularity structure of Yang-Mills theory with
$n_f$ flavors of Majorana adjoint fermions, QCD(adj), for different values
of the fermion mass $m$. We take $1 \le n_f \le 5$, where the upper bound
comes from the requirement of asymptotic freedom, and assume that all of
the quarks have a common mass $m$. We will consider the graded partition
function 
\begin{equation}
\tilde Z(\beta)= \tr [e^{- \beta H} (-1)^F]\,,
\end{equation}
where $H$ is the Hamiltonian and $F$ is fermion number, and $\beta$  is the 
circle size. 
Adjoint QCD is
interesting for many reasons, and has been studied extensively in recent
years, see e.g.~\cite{Kovtun:2007py,Unsal:2007jx,Unsal:2008eg,Unsal:2008ch,Unsal:2007vu, Argyres:2012ka, Anber:2015kea,Anber:2015wha,Anber:2013doa,Anber:2014lba,Bringoltz:2009kb,Bringoltz:2011by,Azeyanagi:2010ne,Cossu:2009sq,Cossu:2013nla,Cossu:2013ora,Hietanen:2009ex,Hietanen:2010fx,Gonzalez-Arroyo:2013bta}.
Crucially, it has a $\mathbb{Z}_N$ center symmetry
\cite{Gross:1980br,Weiss:1981ev}.  The existence of this symmetry makes
the transition from confinement to deconfinement a sharply defined notion.
The deconfinement transition is associated with spontaneous breaking of
$\mathbb{Z}_N$ center symmetry.  

A number of aspects of the $m$ dependence of large $N$ adjoint QCD on
$S^3\times S^1$ was studied by Myers and Hollowood \cite{Hollowood:2009sy},
with a focus on the behavior of center symmetry. Our analysis, which agrees
with theirs on all points of overlap, instead focuses on the behavior of
flow of Hagedorn singularities in the graded partition function. The leading
Hagedorn singularities are associated with an inverse ``temperature''
scale\footnote{We put the word temperature in quotes because we are
discussing the $(-1)^F$-graded partition function, rather than the thermal
partition function.}
$\beta_H$. In general, $\beta_d > \beta_H$ \cite{Aharony:2003sx,Aharony:2005bq},
but  in the $R\Lambda \to 0$ regime on which we focus, $\beta_{d} \to \beta_H$.
So studying the Hagedorn singularity structure of the partition function is
another way to study the fate of color confinement in the theory.

The two features of adjoint QCD of prime importance for our work are
\begin{enumerate}

\item When $m \to \infty$, the fermions decouple, and $\tilde Z(\beta)$ reduces
to the thermal partition function $Z(\beta) = \tr e^{-\beta H}$ of pure
Yang-Mills theory without matter, $n_f = 0$:
\begin{align}
  Z_{\rm YM}(\beta) = \tilde{Z}(\beta,m=\infty) \,.
\end{align}
Then there is a thermal deconfinement transition at $\beta \sim R$ when
$R\Lambda\ll 1$ \cite{Sundborg:1999ue,Polyakov:2001af,Aharony:2003sx}, and
$\beta$ can be interpreted as inverse temperature. 

\item When $m \to 0$, the bosonic and fermionic spectrum of large $N$
adjoint QCD becomes highly correlated in such a way that there are no
phase transition in $\tilde{Z}$ as a function of $\beta$
\cite{Basar:2013sza,Basar:2014jua,Cherman:2018mya}. The compactification
in this limit is non-thermal and $\beta$ is interpreted as circle size. 
\end{enumerate}
 
These two features are related to Hagedorn singularities. To discuss these
singularities, we define the parameter $q=e^{-\beta/R}$. The large $N$ partition
function $Z$ of pure Yang-Mills has infinitely many poles $q_i$ in the domain
$q \in [0, 1]$ \cite{Aharony:2003sx,Polyakov:2001af,Sundborg:1999ue}.  The
inverse Laplace transform of $Z$ is the density of states $\rho(E)$, and each
pole $q_i= e^{-\beta_i /R}$ in $Z$  corresponds to an exponential `Hagedorn'
factor $\rho (E) \sim e^{ \beta_i E} $ in a large-$E$ expansion of $\rho(E)$.
This behavior implies that pure YM theory must have a phase transition to a
deconfined phase at some $\beta \ge \textrm{max}(\beta_i)$.

The behavior of the graded partition function $\tilde{Z}$ of adjoint QCD
with $m=0$ is quite different. To see why, note that $n_f=1$ QCD(adj) is
${\cal N}=1$ supersymmetric (SUSY) Yang-Mills theory, and the supersymmetry
of the theory on $\mathbb{R}^4$ suggests that bosonic and fermionic excitations
should cancel in the $(-1)^F$ graded partition function. Despite the subtlety
that the supersymmetry of pure $\mathcal{N}=1$ SYM is broken by coupling to
the curvature of $S^3$, the remaining cancellations remain strong enough that
$\tilde{Z}(\beta)$ has no singularities for $q\in [0,1]$. The graded partition
function is associated to a graded density of states, $\tilde\rho(E) =
\rho_{\cal B}(E) - \rho_{\cal F}(E)$. Despite  the fact that both $\rho_{\cal B}(E)$
and $\rho_{\cal F}(E)$ exhibit Hagedorn growth, $\tilde{\rho}(E)$ does not.
The absence of Hagedorn behavior in $\tilde\rho(E)$ generalizes to QCD(adj)
with $n_f\geq 2$, implying that there are strong Bose-Fermi cancellations
even without supersymmetry \cite{Basar:2013sza,Basar:2014jua,Cherman:2018mya}. 
In particular, QCD(adj) with $n_f\geq 1$ massless adjoint fermions has no
phase transition on a small $S^3 \times S^1 $ for any $\beta \in [0, \infty)$
\cite{Basar:2014jua,Unsal:2007fb}.  

This leads to an interesting issue which we study in this paper. Consider
turning on a small mass for the adjoint fermion, $m>0$. Since the YM
thermal partition function $Z_{\rm YM}(\beta) = \tilde{Z}(\beta,m=\infty)$
has infinitely many singularities in the physical domain $q \in [0, 1]$,
but $\tilde Z(\beta,m=0)$ does not have any singularities in the physical
domain, the positive real singularities must start to appear at some
finite value $m=m_c$. In this work we study the mechanism by which this
happens. In analogy with the Lee-Yang-Fisher analysis, we show that the
flow of complex Hagedorn singularities $q =e^{- \beta/R + i \mathfrak {t}} \in
{\mathbb C}$ is associated with center-symmetry changing phase transitions
in large $N$ gauge theories on $S^3$. There are two basic scenarios that
lead to the appearance of real singularities, the collision of complex
conjugate singularities, or the inflow of singularities from the endpoint
$q=1$.

%%%%%%%%%%%%%%%%%%%%%%%%%%%%%%%%%%%%%%%%%%%%%%%%%%%%%%%%%%%%%%%%%%%%%%%%%
{\bf  Setting and graded  partition function:}
%%%%%%%%%%%%%%%%%%%%%%%%%%%%%%%%%%%%%%%%%%%%%%%%%%%%%%%%%%%%%%%%%%%%%%%%%
We now describe the setting for our analysis in more detail. When Yang-Mills
theories are placed on a spatial $S^3$, the gauge fields and fermions pick
up effective masses due to the curvature of $S^3$ \cite{Aharony:2003sx}.
When $R\Lambda \ll 1$, the curvature is large, so the effective coupling
$\lambda(1/R)$ becomes small, and we can integrate out all of the non-zero
modes of the gauge fields at one loop, which amounts to working at $\lambda
= 0$.  But the part of the path integral associated with the holonomy 
\be
\alpha := \frac{1}{{\rm Vol}({\cal M})} \int_{S^3 \times S^1} \, A_0,
\ee
must be treated exactly, and the  path integral reduces to a matrix
integral over $\alpha$. One can diagonalize $\alpha$ using a global gauge
transformation,
\be
\alpha_{\rm diag} = {\beta} ^{-1} {\rm diag}(\theta_1,\cdots,\theta_N).
\ee
and then 
\begin{align}
\tilde{Z} = \int D\alpha_{\rm diag} e^{-S_{\rm eff}[\alpha_{\rm diag}]}
\end{align}
where $S_{\rm eff}[\alpha_{\rm diag}]$ includes a contribution from the Haar
measure. The calculation of the one-loop effective action  $S_{\rm eff}$ is
described in \cite{Aharony:2003sx,Hollowood:2009sy}. At large $N$, the
difference between $SU(N)$ and $U(N)$ gauge groups becomes negligible for
the purposes of this paper, but the effective action of the $U(N)$ theory
is simpler. So we will work with $S_{\rm eff}$ for $U(N)$ gauge theory.
On small $S^3$, this effective action takes the form
\cite{Aharony:2003sx,Hollowood:2009sy}
\be
 S_{\rm eff}(\alpha) &=& \sum_{n=1}^{\infty} 
   \frac{ f_{n}(q, m)}{n}
 |\Tr (P^n)|^2 \, , 
\label{eq:S_alpha}
\ee
where  $P=e^{i{\beta} \alpha}$ is the Polyakov line, and
\beq
f_{n} (q^n, mR):= 1-z_v(q^n) +   n_f  z_f(q^n, m R),
\label{eq:fn_majorana}
\eeq
Here, $z_v$ and $z_f$ are single particle partition functions for vector
fields and fermions, and 
\begin{align}
z_v(q) &= 2 \sum_{\ell=1}^{\infty} \ell (\ell+2) e^{-{\beta} (\ell+1)/R} 
   = \frac{6 q^2 - 2  q^3 }{(1-q)^3}, 
\end{align}
When $m=0$, $z_f$ takes the simple form
\begin{align}
z_f(q) &= 2 \sum_{\ell=1}^{\infty} \ell (\ell+1) e^{-{\beta} (\ell+1/2)/R}  
  =  \frac{4q^{3/2}}{(1-q)^3}.
\end{align}
But when $m R >0$, it is simpler to write $z_f$ as a function of
${\beta} /R = - \log q$ in the form \cite{Hollowood:2009sy}
\be
&&z_f({\beta} /R,m_fR) = 2 \sum_{\ell=1}^{\infty} \ell (\ell+1) 
   e^{-{\beta}  \sqrt{(\ell+1/2)^{2}+m_f^2 R^2}/R} \nl
&&= \frac{2m_f^2 R^3}{{\beta} }K_2(m_f{\beta} )-\frac{m_f R}{2}K_1(m_f{\beta} ) \nl
&& \mbox{}+ 4 \int^{\infty}_{m_fR} dx \, 
  \frac{x^2 + 1/4}{e^{2 \pi x} + 1} \sin({\beta}  \sqrt{x^2-m_f^2 R^2}/R^2).
\label{eq:zf_sum} \qquad 
\ee
It is easy to verify that when $q \to 0 \; (\beta \rightarrow \infty)$ , all
of the coefficients $f_{n}$ are positive.  This implies that $S_{\rm eff}$
is minimized if all moments $\rho_n= \Tr (P^n)$($n\geq 1$) are zero, and
the theory is in the center-symmetric (that is, confining) phase. The
partition function is obtained by integrating over the Gaussian modes
$\rho_n$, which gives 
\be
 \tilde{Z}(q) =   \prod_{n=1}^{\infty} \frac{1}{ f_{ n} (q^n, mR)}\, . 
\ee
This expression is physically valid so long as all $f_n$ are positive,
which is in general only true for some range of values of $q$ in the real
interval $[0, q_c], q_c \le 1$, with $q_c = q_c(m)$.  However, we find it
useful to analytically continue $\tilde{Z}$ to a function of $q$ within
the unit disk $|q| <1$ in the complex $q$ plane, and in the rest of this
paper $\tilde{Z}$ will denote this analytically-continued quantity.

It will be useful to compare the graded partition function $\tilde{Z}$
to the standard thermal partition function $Z$. On  small $S^3 \times S^1$,
the thermal partition function is obtained by replacing 
\beq
 f_{n} (q^n, mR) \to b_n(q^n, mR)
\eeq
where 
\begin{align}
b_n(q^n, mR) =   1-z_v(q^n) + (-1)^n n_f  z_f(q^n, m R)\,,
\end{align}
so that 
\be
Z =   \prod_{n=1}^{\infty} \frac{1}{ b_{ n} (q^n, mR)}\, . 
\ee
It is known since Ref.~\cite{Aharony:2003sx,Hollowood:2009sy} that $Z$
has infinitely many poles in the interval $q \in [0,1]$ for any $n_f \ge 0$,
corresponding to real positive solutions of $b_{n} (q^n, mR)= 0$ for
$n\geq 1$ in this range. As discussed in \cite{Basar:2014hda,Basar:2014jua},
there are also infinitely many complex poles within the unit disc $|q|\leq 1$,
where $q \in \mathbb C$, but for the Hagedorn growth of the  spectral density
only the poles $q \in [0,1]$ are relevant.

{\bf Hagedorn transseries:} To understand the connection between singularities
in the partition function and Hagedorn behavior of the density of states, let
us write the $(-1)^F$ density of states $\tilde{\rho}(E) = \rho_{\rm bosonic}(E)
-  \rho_{\rm fermionic}(E) $ as a transseries expansion in variable $x = 1/(ER)$
around the point $x=0$. A natural ansatz for the form of $\tilde{\rho}(E)$ is
\begin{align}
R^{-1}\tilde{\rho}(E) &= e^{\beta_1/x} \sum_{n=0}^{\infty}  a_{1,n} x^{n}
   + e^{\beta_2/x} \sum_{n=0}^{\infty} a_{2,n} x^{n} + \cdots
\nonumber \\
 & +  \frac{1}{2}(e^{\gamma_1/x}+e^{\gamma^{*}_{1}/x})
        \sum_{n=0}^{\infty} c_{1,n} x^{n}
\nonumber \\
 & +  \frac{i}{2}(e^{\delta_{1}/x}-e^{\delta^{*}_{1}/x})
        \sum_{n=0}^{\infty} d_{1,n} x^{n} + \cdots
        \label{general}
\end{align}
where $a_{n,k},c_{n,k}, d_{n,k}, \beta_n, \gamma_n, \delta_n$ are dimensionless
parameters, $\beta_{n}\in\mathbb{R}$ with $\beta_1 > \beta_2 > \cdots$,
and $\gamma_{n} \in \mathbb{\mathbb{C}}$ with $\textrm{Re}\gamma_{1} >
\textrm{Re}\gamma_{2} > \cdots$ and $\textrm{Im}\gamma_n \neq 0$ for all
$n$ (and similarly for $\delta_n$). The inverse Hagedorn temperature scales
are $\beta_{H, n} = \beta_n R$, which are physical when $\beta_n>0$.  

This is only the simplest
ansatz for $\tilde{\rho}(E)$ which incorporates the phenomena we want to
describe. Physically, however, we expect the transseries representation of
$\tilde{\rho}(E)$ to depend on more transmononomials of $x$, such as $\log(x)$
and $x^{b}$ with $b \notin \mathbb{N}$.  Determining the complete transseries
representation of $\rho(E)$ is a currently open problem even in the simplest
cases such as $m=0$ and $m =\infty$, where $\tilde{Z}$ (and also $Z$) can be
written as products of elliptic functions \cite{Basar:2015xda,Basar:2015asd}.
It is likely that this problem can be solved using the methods of
Ref.~\cite{COSTIN20091370}, and it would be interesting to do so in future
work.

To see how this ansatz for the density of states relates to the partition
function, note that we can write 
\begin{align}
\tilde{Z}(\beta) = \tr (-1)^F e^{-\beta H} = \int_{E_0}^{\infty} d E\,
     \tilde{\rho}(E) e^{-\beta E} \,.
\end{align}
where $E_0 = c/R$ with $c = 3/2$. The cut off on the energy integral at
$E_0$ takes into account that the spectrum of 4d adjoint $U(N)$ QCD on
small $S^3$  is gapped with a gap that interpolates between $E_0$ at
$m=0$ and $4/3 E_0$ at $m = \infty$.  So $\tilde{\rho}(E)$ must always
vanish for $E < E_0$.   Evaluating the integral gives
\begin{align}
  \tilde{Z}(\beta) &= \frac{1}{\beta - \beta_1}
  \biggg\{ a_{1,0} +  \bigg[c a_{1,0}
      -  a_{1,1}\left( \gamma_{E}
            +\log[c(\beta - \beta_{1}) ] \right)  
\nonumber\\
 & +\sum_{n=2}^{\infty} \frac{a_{1,n}}{c(n-1)}   \bigg]
     \left(\beta-\beta_{1}\right) \nonumber\\
 & +  \bigg[\frac{c^2}{2} a_{1,0}  + c a_{1,1}
       + a_{1,2}\left( \gamma_{E} - 1
       + \log[c(\beta - \beta_{1}) ] \right) \nonumber\\
 & +\sum_{n=3}^{\infty} \frac{a_n}{c(n-1)}   \bigg]
     \left(\beta-\beta_{1}\right)^2\biggg\}
 + \ldots \nonumber \\
 & + \frac{1}{\beta - \beta_2}
    \bigg[a_{2,0}+ \ldots\bigg] + \ldots
\end{align}
This expression illustrates that exponentially-growing terms in
$\tilde{\rho}(E)$ generate singularities in $\tilde{Z}(\beta)$.  The
coefficients $\beta_{n}, a_{n,k}, \gamma_n, c_{n,k},\delta_n, d_{n,k}$ all
depend on the quark mass $m$.  

We note that the thermal partition function $Z(\beta)$ and the thermal
(not graded) density of states $\rho(E)$ can be written in the same
form, but of course the numerical values of the parameters entering
the thermal density of states transseries are different from the
$(-1)^F$-graded one.

%%%%%%%%%%%%%%%%%%%%%%%%%%%%%%%%%%%%%%%%%%%%%%%%%%%%%%%%%%%%%%%%%%%%%%%%%%
\begin{figure}[t]
\begin{center}
\includegraphics[clip, width=0.75\hsize]{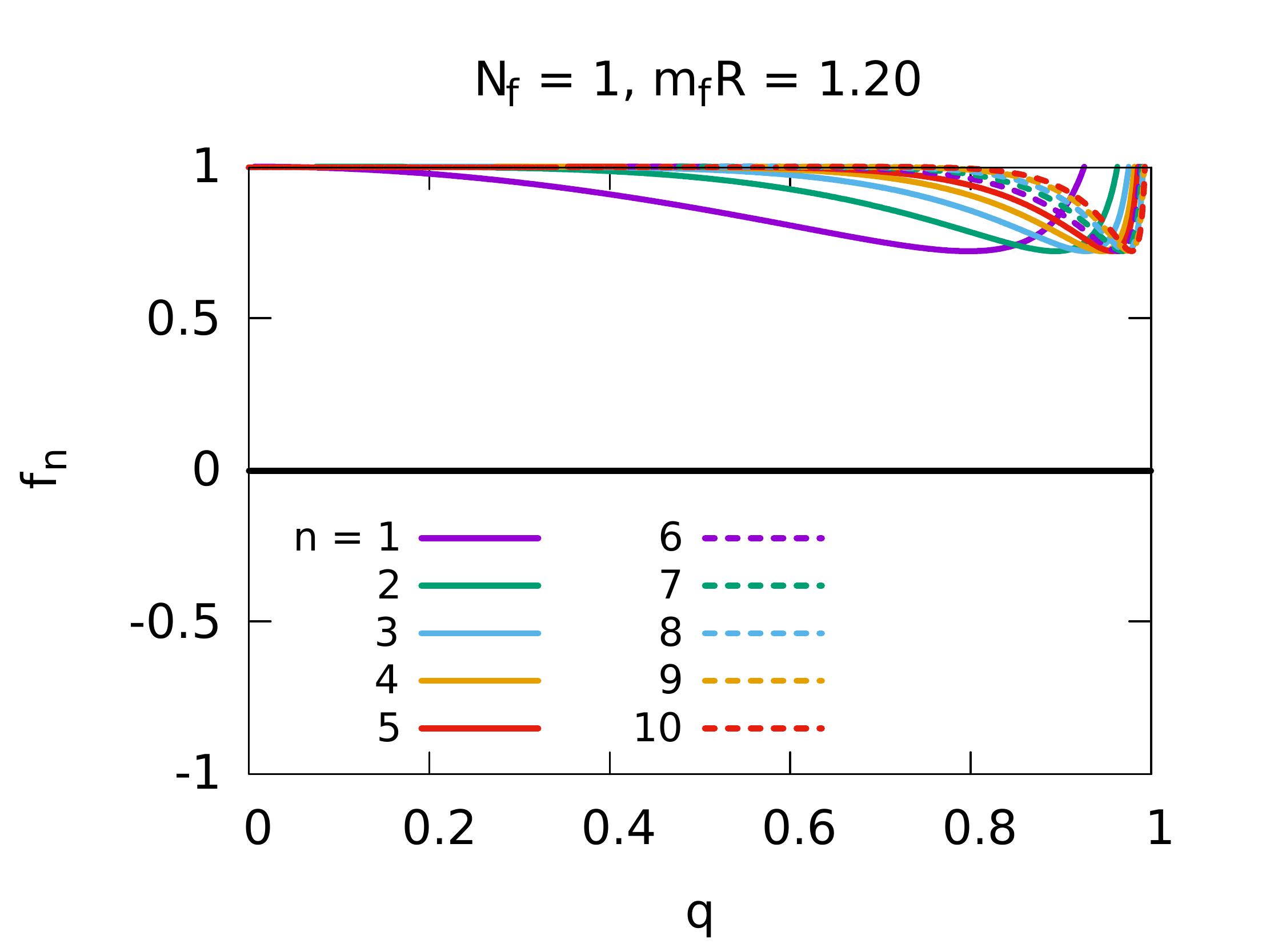}
\end{center}
\begin{center}
\includegraphics[clip, width=0.75\hsize]{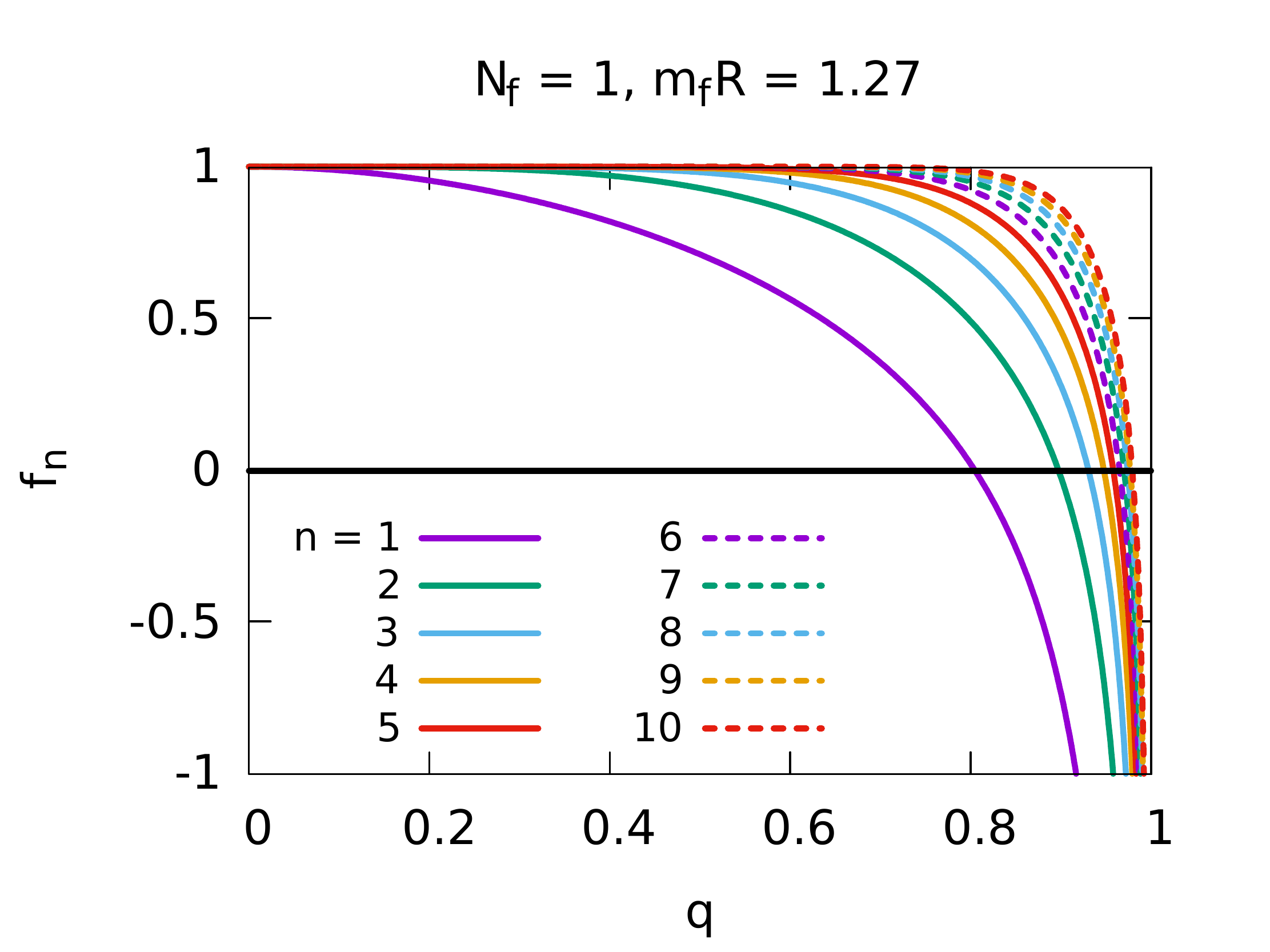}
\end{center}
\caption{ \label{fig:fn0_Nf1}
Plots of $f_{n}(q)$ as a function of real $q\in [0,1]$ for $n_f=1$. 
The top panel shows $m_fR=1.20$, below the critical mass. The 
bottom panel shows $m_fR=1.27$, above the critical mass.}
\end{figure}
%%%%%%%%%%%%%%%%%%%%%%%%%%%%%%%%%%%%%%%%%%%%%%%%%%%%%%%%%%%%%%%%%%%%%%%%%%

%%%%%%%%%%%%%%%%%%%%%%%%%%%%%%%%%%%%%%%%%%%%%%%%%%%%%%%%%%%%%%%%%%%%%%%%%%
\begin{figure}[t]
\begin{center}
\includegraphics[clip, width=0.75\hsize]{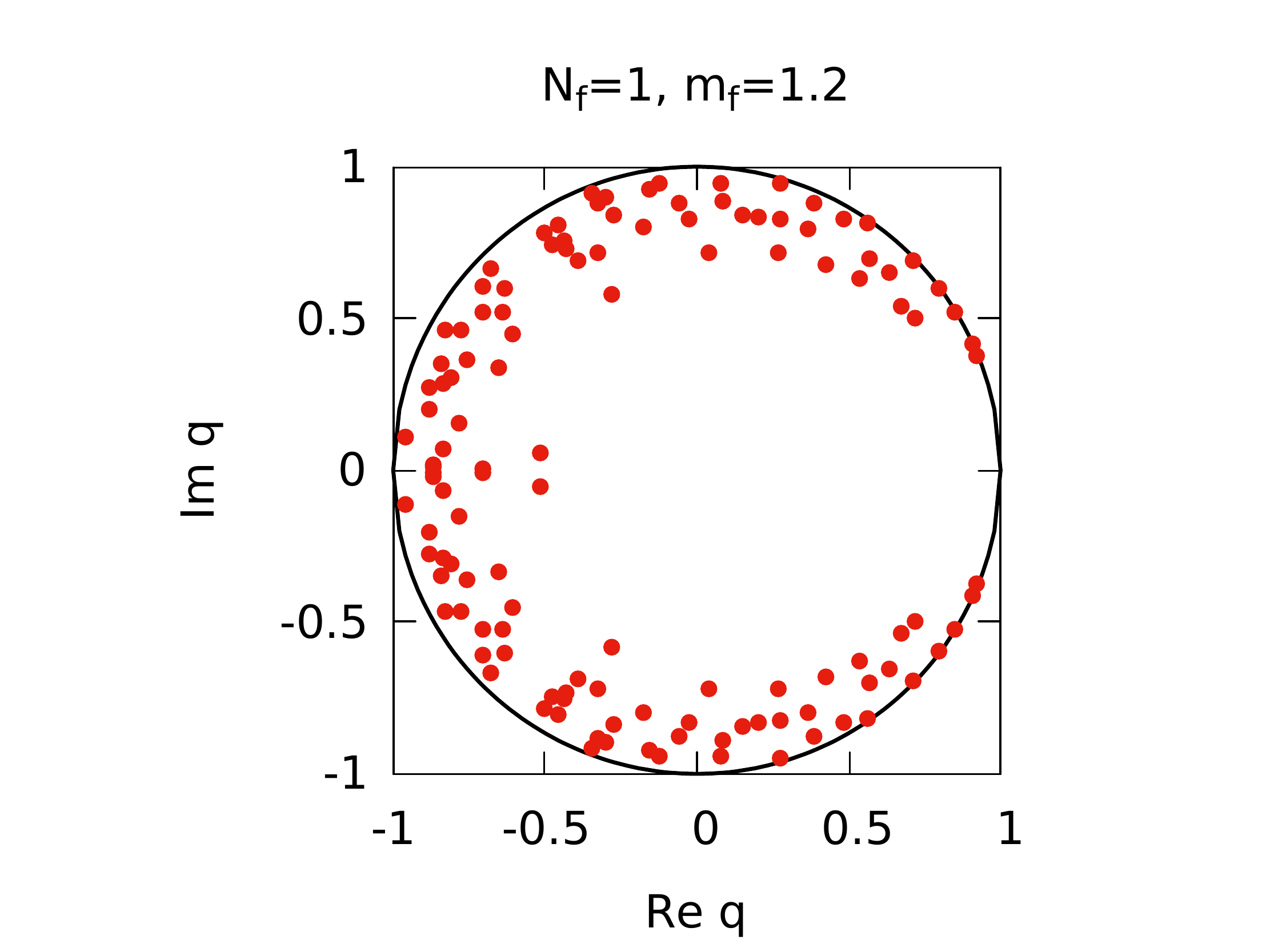}
\end{center}
\begin{center}
\includegraphics[clip, width=0.75\hsize]{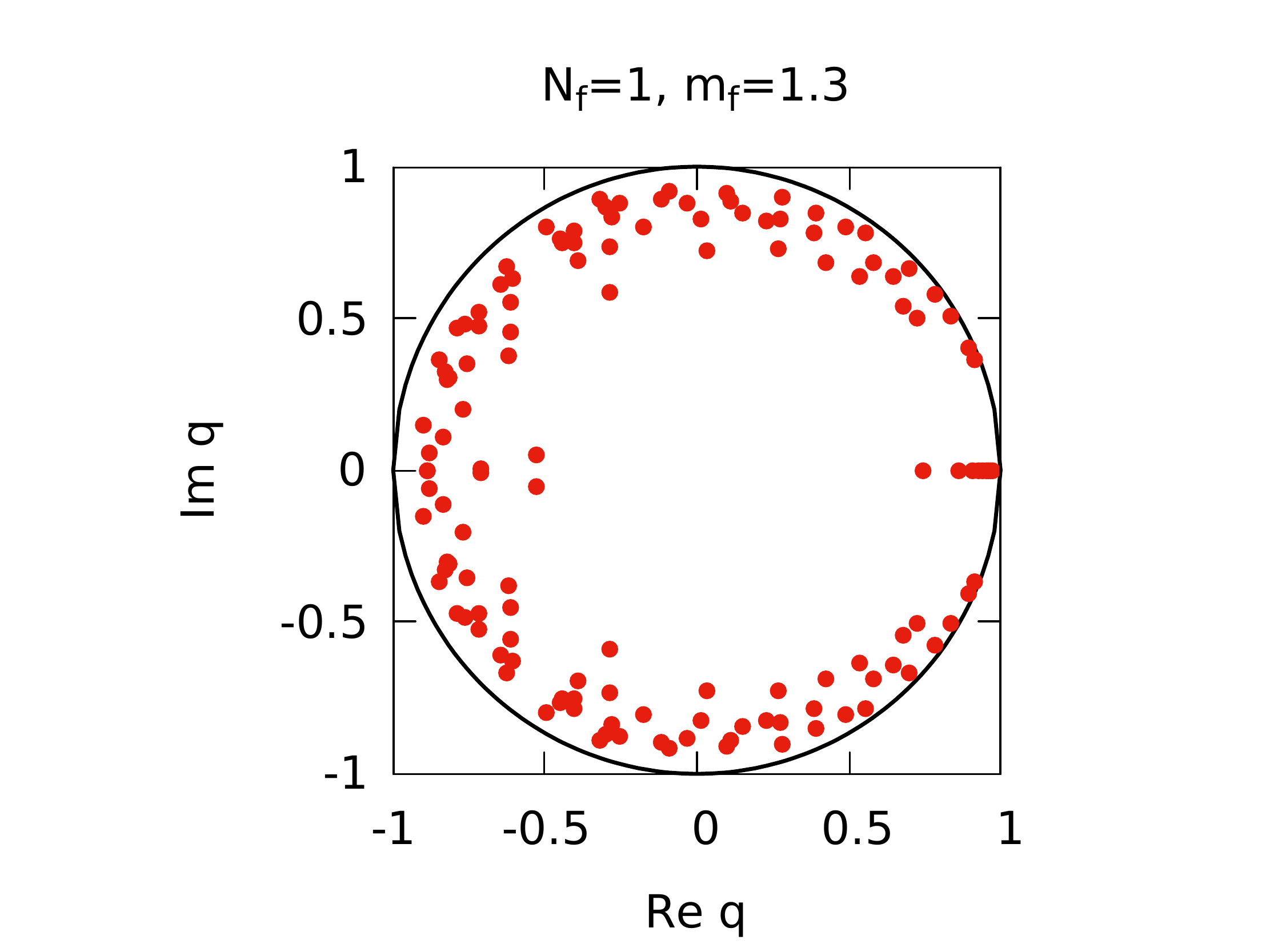}
\end{center}
\caption{\label{fig:q_plane_nf1}
Complex $q$-plane for $N_f=1$ near $m_f=m_f^*$. Top panel:
$N_f=1,m_fR=1.2$. Bottom panel: $N_f=1,m_fR=1.3$.}
\end{figure}
%%%%%%%%%%%%%%%%%%%%%%%%%%%%%%%%%%%%%%%%%%%%%%%%%%%%%%%%%%%%%%%%%%%%%%%%%%

%%%%%%%%%%%%%%%%%%%%%%%%%%%%%%%%%%%%%%%%%%%%%%%%%%%%%%%%%%%%%%%%%%%%%%%%%%
{\bf Flow of Hagedorn singularities:}
%%%%%%%%%%%%%%%%%%%%%%%%%%%%%%%%%%%%%%%%%%%%%%%%%%%%%%%%%%%%%%%%%%%%%%%%%%
Let us now work out the flow of the Hagedorn parameters $\beta_{n}$
and $\gamma_n$ as function of the quark mass $m$. One can easily
verify that $f_{n} (q^n, m = 0)$ is positive for all $q \in [0,1]$, so there
are no poles of $\tilde{Z}(q)  $ for $q \in [0,1]$ \cite{Basar:2014jua}.
Correspondingly, there is no Hagedorn growth in the graded density of
states, $\tilde{\rho} (E) $. This means that $\beta_{n} < 0$ for
all $n$ at $m=0$, but it does not constrain $\gamma_n$.  Physically,
these remarks imply that the infinitely large number of Hagedorn-growing
terms in both $ \rho_{\cal B}(E)$ and $\rho_{\cal F}(E)$ cancel against each
other, leaving a function with sub-exponential growth in $E$.  The main
point of \cite{Basar:2013sza,Cherman:2018mya} is that this is not an
artifact of working at small $R\Lambda$.  It is related to the fact that
it enjoys large $N$ volume independence~\cite{Kovtun:2007py,Bringoltz:2009kb,Azeyanagi:2010ne,Catterall:2010gx,Hietanen:2009ex,Hietanen:2010fx,Bringoltz:2011by,Gonzalez-Arroyo:2013bta,Lohmayer:2013spa,Perez:2015yna}.
In infinite volume, on $\mathbb R^3 \times S^1$ and   $N=\infty$,  adjoint
QCD  endowed with periodic boundary conditions does not have any phase
transitions. This implies  that the physical gauge-invariant bosonic and
fermionic excitations of adjoint QCD are very tightly correlated despite
the manifest absence of supersymmetry in adjoint QCD for $n_f>1$.
 
Now suppose $m \neq0$.  As mentioned in the introduction, at $m= \infty$
the fermions decouple from the spectrum of the theory, so that 
\be
{\lim}_{m \rightarrow \infty}   \tilde{Z}(q, m)   =  Z_{\rm YM}(q)
\ee
Therefore, $\tilde{Z} (q, 0)$ is free of poles $q \in [0,1]$, but 
$\tilde{Z}(q,\infty)$ has infinitely many poles.  So when $m \to \infty$,
$\beta_1$ along with an infinite subset of the $\beta_{n}$
coefficients are positive.   In the free limit, the Hagedorn temperature
$\beta_1$ corresponds to a center-symmetry changing phase transition.  

Combining these observations, we see that as $m$ is varied in $[0, \infty)$
a pole on the real interval $[0,1]$ has to appear at some critical mass $m_c$.
Tracking the motion of the Hagedorn singularities is a difficult task in
general, but at small $R\Lambda$ it can be done explicitly thanks to the
fact that the theory is weakly coupled.  

 We find two distinct types of Hagedorn singularity flow.

{ \bf Adjoint QCD with $\mathbf{n_f = 1}$:}  
This is the only value of $n_f$ at which the $m=0$ limit of the theory
has $\mathcal{N}=1$ supersymmetry on $\mathbb{R}^3 \times S^1$.  Perhaps
not coincidentally, we find that the flow of the Hagedorn singularities
at $n_f=1$ is quite different from the flow at higher $n_f$. For $n_f = 1$,
we show the functions $f_{n}(q)$ along the real interval $q\in[0,1]$ in
Fig.~\ref{fig:fn0_Nf1}, while the singularities in the complex $q$-plane
are shown in Fig.~\ref{fig:q_plane_nf1}. Both plots show the behavior
just below and just above the critical mass $ m_c R\simeq 1.22 \,$.

One can see that as  $m$ increases pass $m_c$, all $f_{n}(q)$ change sign
for some value of $q$ and stay negative as $q \rightarrow 1$. A related
observation is that the zeros in $f_n(q)$ appear from $q=1$ limit as $m$ is
increased.  This is reflected in the complex $q$-plane, where singularities
of $\tilde{Z}$ flow in from the point $q=1$ once $m>m_c$. The distribution
of these singularities approaches the distribution of the singularities
of the partition function of pure YM theory (obtained in the limit $m\to
\infty$) by moving inward along the real axis.   The singularities on
$q \in [0,1)$ have an interpretation as terms that scale as $e^{ {\beta}_{n}E}$
with $\beta_n>0$, in the graded density of states.  

For $m<m_c$, the theory preserves the full ${\mathbb Z}_N$ center-symmetry
at any value of ${\beta} /R$. For $m>m_c$,  the ${\mathbb Z}_N$ center-symmetry
is broken completely --- that is, it is broken to ${\mathbb Z}_1$.  However,
as $m_c$ is lowered from $\infty$ to $m_c$, the  center-breaking scale of
one-flavor adjoint QCD increases smoothly from the deconfinement temperature
of pure YM theory, $T_{\rm YM}$,  to $\beta^{-1} = \infty$, connecting a thermal
phase transition to a quantum phase transition.

%%%%%%%%%%%%%%%%%%%%%%%%%%%%%%%%%%%%%%%%%%%%%%%%%%%%%%%%%%%%%%%%%%%%%%%%%%
\begin{figure}[t]   
\begin{center}
\includegraphics[clip, width=0.75\hsize]{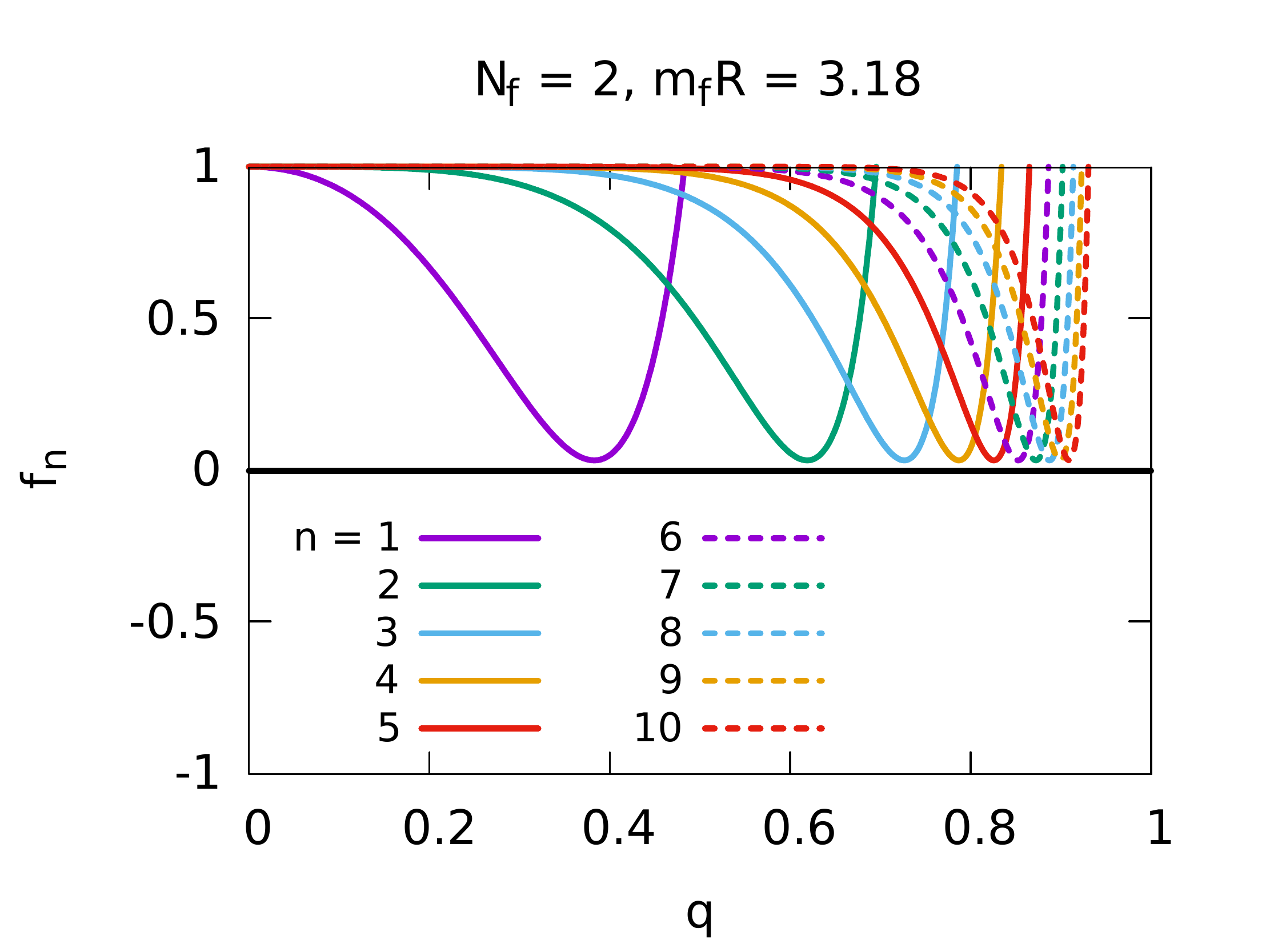}
\end{center}
\begin{center}
\includegraphics[clip, width=0.75\hsize]{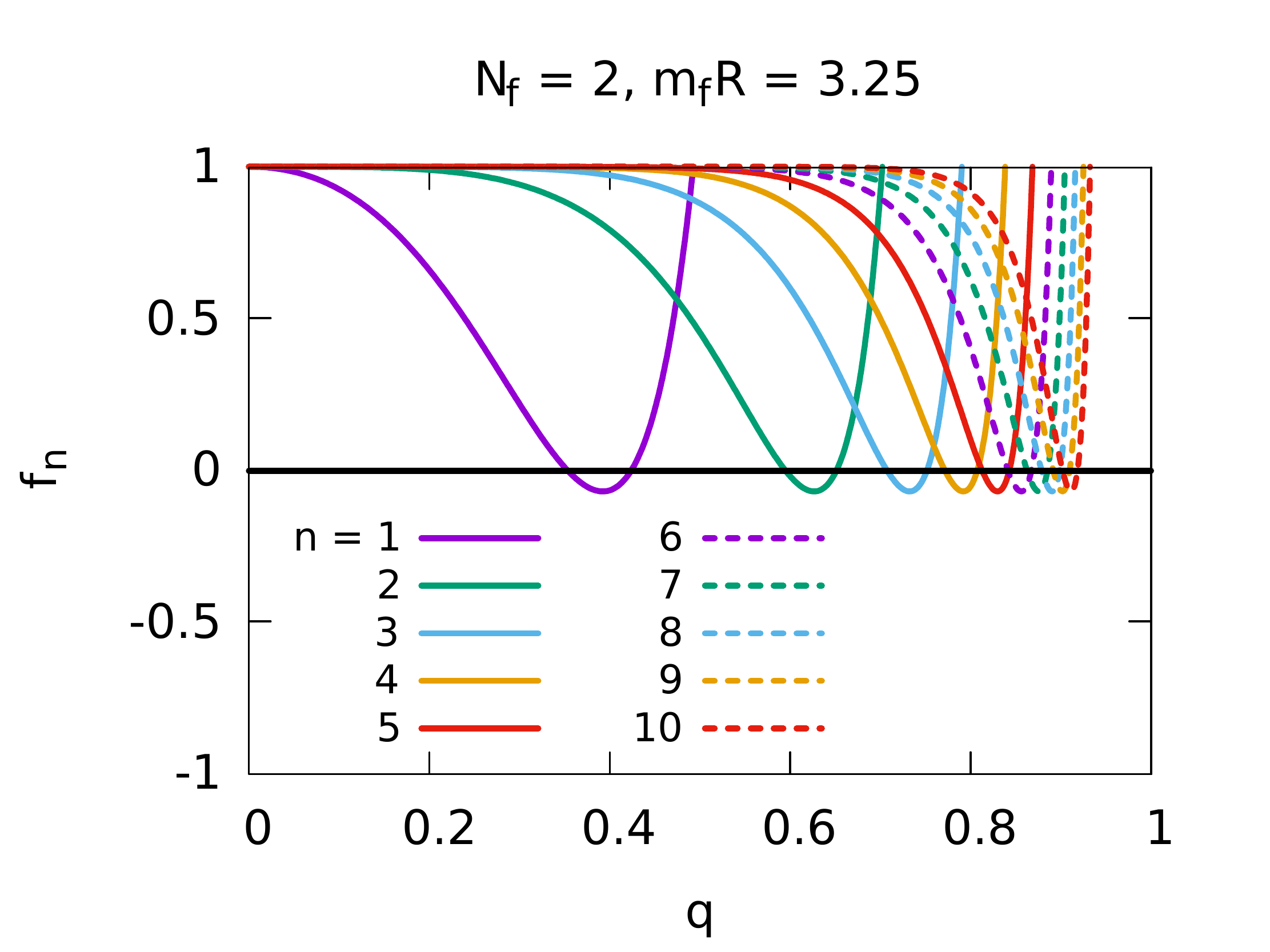}
\end{center}  
\caption{ \label{fig:fn0_Nf2}
Plots of $f_{n}(q)$ as a function of real $q\in [0,1]$ for $n_f=2$. 
Top panel: $m_fR=3.18$. Bottom panel: $m_fR=3.25$.}
\end{figure}
%%%%%%%%%%%%%%%%%%%%%%%%%%%%%%%%%%%%%%%%%%%%%%%%%%%%%%%%%%%%%%%%%%%%%%%%%%

%%%%%%%%%%%%%%%%%%%%%%%%%%%%%%%%%%%%%%%%%%%%%%%%%%%%%%%%%%%%%%%%%%%%%%%%%%
\begin{figure}[t]
\begin{center}
\includegraphics[clip, width=0.75\hsize]{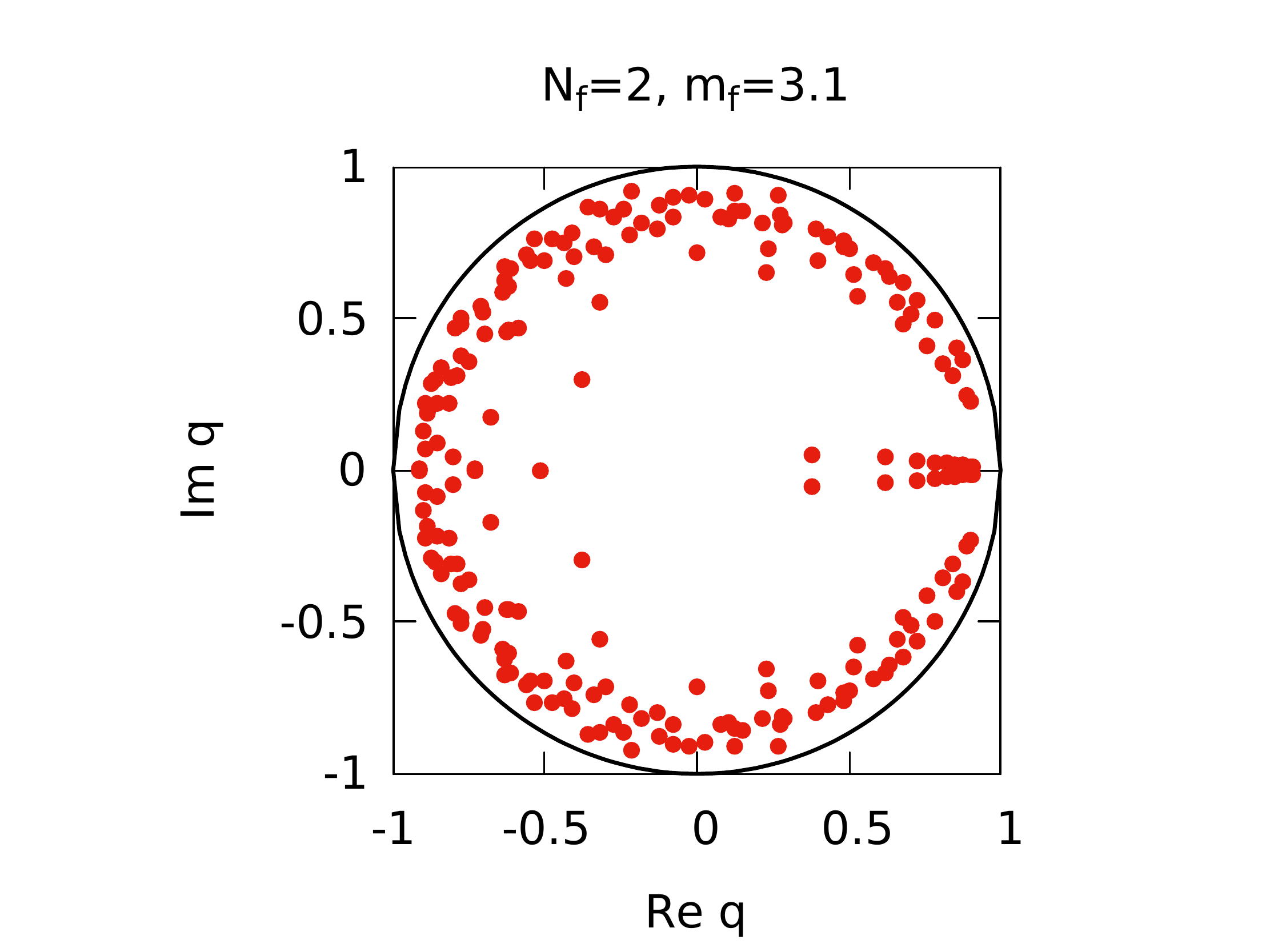}
\end{center}
\begin{center}
\includegraphics[clip, width=0.75\hsize]{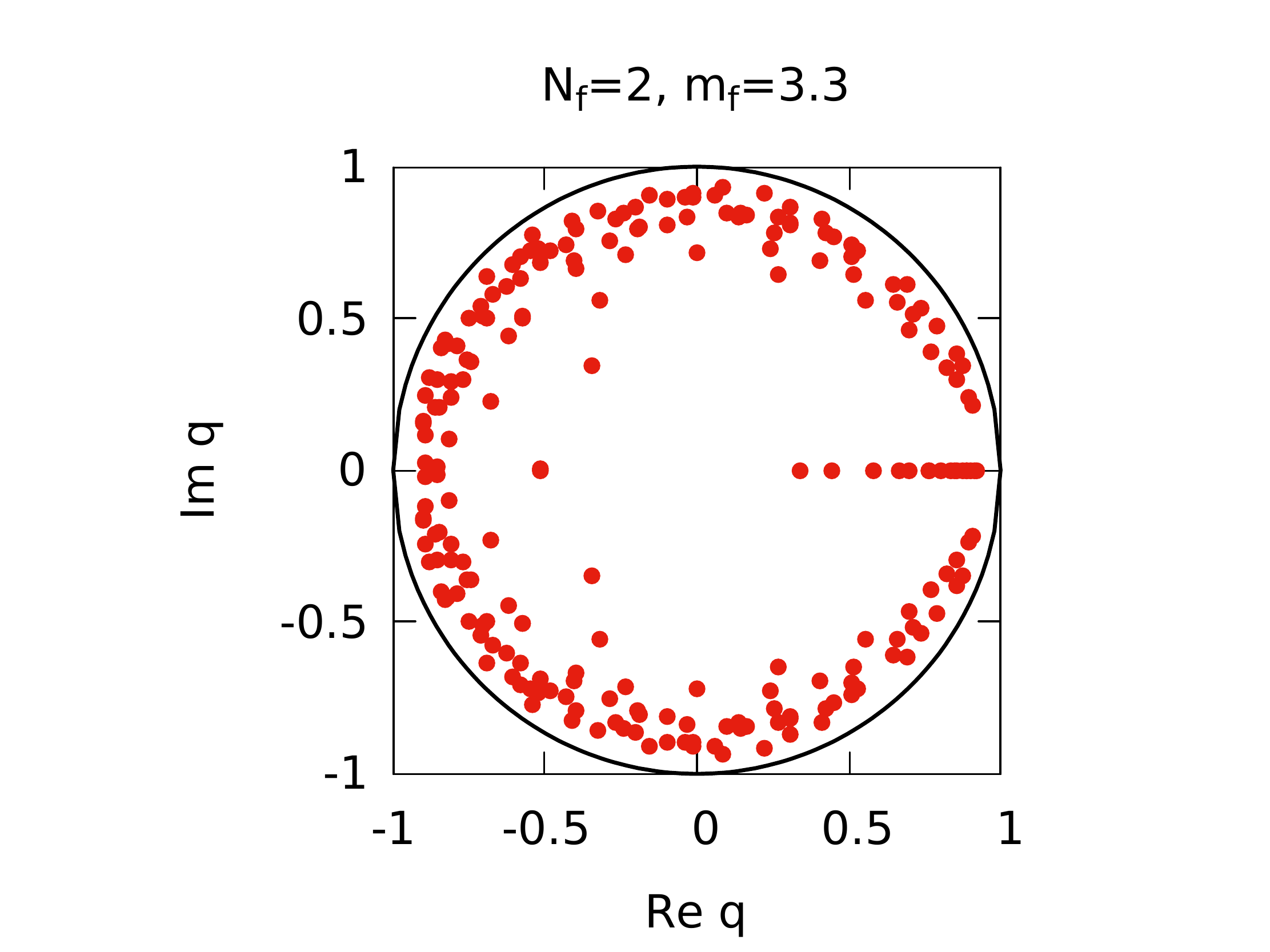}
\end{center}
\caption{\label{fig:q_plane_nf2}
Complex $q$-plane for $N_f=2$ near $m_f=m_f^*$. Top panel:
$N_f=2, m_fR=3.1$.  Bottom panel: $N_f=2,m_fR=3.3$.}
\end{figure}
%%%%%%%%%%%%%%%%%%%%%%%%%%%%%%%%%%%%%%%%%%%%%%%%%%%%%%%%%%%%%%%%%%%%%%%%%%

{\bf Adjoint QCD with $\mathbf{n_f \ge 2}$:} We have checked that the
behavior for $n_f = 2$ and higher $n_f$ is qualitatively identical, so for
simplicity we focus our discussion on the case $n_f =2$.  The function
$f_{n}(q)$ and the singularities in the complex $q$-plane are shown in
Fig.~\ref{fig:fn0_Nf2} and~\ref{fig:q_plane_nf2}. The critical value of
$m$ is now $m_cR \simeq 3.20$.

When $m < m_c$  the functions $f_{n}(q)$ do not cross the $f_{n}=0$ line,
but have minima that approach it as $m \to m_c$. This is reflected by
complex conjugate pairs of singularities in the $q$-plane that approach,
but do not cross, the positive real axis. The theory preserve the full 
${\mathbb Z}_N$ center-symmetry at any value of ${\beta} /R$ when $m < m_c$. 
However, when $m>m_c $, the pairs of singularities collide and move onto
the positive real $q$  axis.   This corresponds to a deconfining phase
transition, which we now examine. 

The first big difference one can see from the plots in
Figs.~\ref{fig:fn0_Nf2},\ref{fig:q_plane_nf2} is that the $\mathbb Z_N$
changing phase transition scale is finite if $m$ approaches $m_c$ from
above. This should be contrasted with what we saw when $n_f = 1$, when
the   $\mathbb Z_N$  changing transition scale $\beta \rightarrow 0$ 
when $m$ approaches $m_c$ from above.  

Next, let us understand the properties of the transition in more detail.
Suppose that $m\gg m_c$ (large mass limit)  and lower ${\beta} /R$ starting
from ${\beta} /R=\infty$ \cite{Hollowood:2009sy}. The theory starts out in
the center-symmetric phase. At a critical value of ${\beta} /R$ there is a
phase transition to a phase where $\mathbb Z_N \rightarrow \mathbb Z_1$. As
${\beta} /R$ is reduced further, $f_1$ eventually becomes positive, while
$f_{n \geq 2}$ are still negative. Hence, the center partially restores to
$ \mathbb Z_2$. In fact, as ${\beta} /R$ is reduced, in the large $N$ limit
we observe the pattern 
\be
\mathbb Z_N \rightarrow  \mathbb Z_1  \rightarrow \mathbb Z_2  
             \rightarrow \mathbb Z_3  \rightarrow \mathbb Z_4  \ldots \, .  
\label{pattern1}
\ee
In the phase where the center-symmetry breaking pattern is $\mathbb Z_N 
\rightarrow \mathbb Z_k$, the gauge structure becomes $ [U(N/k) ]^k $ where
eigenvalues of Wilson line split into $k$ bunches each of which has $N/k$
coincident eigenvalues.   

An interesting structure appears for $(m -m_c)R= \epsilon$ small and positive.
Reducing ${\beta} /R$, we observe center-symmetry breaking $\mathbb Z_N
\rightarrow \mathbb Z_1$ at the location where $f_1=0$. But $f_1$ turns
around and becomes positive again while all other $f_{n\ge 2}$ are still
positive. Continuing to decrease ${\beta} $, $f_2$ switches sign, and
repeats the same pattern. Hence, there are ``pockets'' of restored
$\mathbb Z_N$ symmetry, with the pattern 
\be
\mathbb Z_N \rightarrow \mathbb Z_1  \rightarrow \mathbb Z_N  
            \rightarrow \mathbb Z_2  \rightarrow \mathbb Z_N 
            \rightarrow \mathbb Z_3  \ldots \, . 
\label{pattern2}
 \ee
As $\epsilon$ is increased, these pockets of full-center restoration close, 
and the pattern turns into that of Eq.~\eqref{pattern1}. 

%%%%%%%%%%%%%%%%%%%%%%%%%%%%%%%%%%%%%%%%%%%%%%%%%%%%%%%%%%%%%%%%%%%%%%%%%%%%%
{ \bf Phase transition and graded density of states:}
%%%%%%%%%%%%%%%%%%%%%%%%%%%%%%%%%%%%%%%%%%%%%%%%%%%%%%%%%%%%%%%%%%%%%%%%%%%%%
In the trans-series discussion, we provided a general form for the Hagedorn
transseries \eqref{general}. It is instructive to study the phase transition
in terms of the graded density of states $\tilde \rho(E)$, and to investigate
how the transseries structure for  $\tilde \rho(E)$ changes as a function
of mass parameter.  We consider the case $n_f\geq 2$. As $m \rightarrow
m_c$, infinitely many complex conjugate poles coalesce on the real axis 
$q \in [0,1]$. For $m = m_c+ 0^{+}$, we can enumerate the poles on the 
positive real axis $\mathbb R^{+}_{\beta}$ as $\beta_{H_1} > \beta_{H_2} >
\beta_{H_3} >  \ldots$, where  $e^{\beta_{H_1} E}$ is the leading Hagedorn growth
in the density of states. To get a feeling for the physics, let us study a
simplified toy model where the partition function only has the two most
dominant poles at $\beta = \beta_{H_1}, \beta_{H_2}$.

 The density of states is the inverse Laplace transform of the partition
function. A toy model describing the collision of the two poles is defined by 
\begin{align} 
\tilde Z(\beta, m) = \left\{ \begin{array}{ll}  
(\beta - \beta_{H_1})^{-1}(\beta - \beta_{H_2})^{-1}  & \; m> m_c \cr
(\beta - \beta_{H_1})^{-2}  &  \; m= m_c \cr
(\beta - \beta_{H})^{-1}(\beta - \beta_{H}^*)^{-1}  & \; m< m_c
\end{array} \right.
\end{align}
where $\beta_{H}=\beta_{H_R} +i \beta_{H_I} \in \mathbb C$. When $m$ is large,
there are two isolated real poles at $\beta = \beta_{H_1}$ and $\beta_{H_2} $.
But when $m$ is decreased towards $m=m_c$ the two isolated real poles merge
into a double pole, and move into the complex plane as complex conjugate
poles for $m< m_c$. The inverse Laplace transform is 
\begin{align} 
\tilde \rho(E) = \left\{ \begin{array}{ll}  
\frac{ e^{\beta_{H_1} E} - e^{\beta_{H_2} E}}{ \beta_{H_1} - \beta_{H_2}}
  & \; m> m_c \cr 
E\, e^{\beta_{H_1} E}   &  \; m= m_c \cr 
 \beta_{H_I}^{-1}\, e^{\beta_{H_R} E} \sin(\beta_{H_I} E)   & \; m< m_c
\end{array} \right.
\end{align}
In the regime where the center symmetry is completely broken, the Hagedorn
growth $e^{\beta_{H_1} E}$ is the dominant contribution to $\tilde \rho(E)$. In
the domain where center is intact, $\tilde \rho(E)$ is oscillatory, consistent
with the general transseries structure \eqref{general}.  The frequency
of oscillation is controlled by the imaginary part of the Hagedorn temperature,
and in the limit where $\beta_{H_I} \rightarrow 0$, the oscillations disappear.
Oscillations of the spectral density in the limit $m\rightarrow 0$ were 
studied in \cite{Basar:2014jua}. These authors noted that Bose-Fermi 
cancellations are responsible for the disappearance of the Hagedorn growth
in the graded density of states. This cancellation, unlike supersymmetry, 
does not take place level-by-level. Instead, the cancellations involve many
neighboring levels, in a pattern essentially identical to that of
``misaligned supersymmetry'' in string theory \cite{Kutasov:1990sv,Dienes:1994np,Dienes:1995pm}. Eventually, once $m>m_c$, the Bose-Fermi balance breaks
down and Hagedorn growth emerges.

%%%%%%%%%%%%%%%%%%%%%%%%%%%%%%%%%%%%%%%%%%%%%%%%%%%%%%%%%%%%%%%%%%%%%%%%%%
{\bf Outlook:}
%%%%%%%%%%%%%%%%%%%%%%%%%%%%%%%%%%%%%%%%%%%%%%%%%%%%%%%%%%%%%%%%%%%%%%%%%%
In this work we have studied to motion of Hagedorn singularities in the
complexified $q=\exp(-{\beta} /R)$ plane of large $N$ Yang-Mills theory
with $n_f$ adjoint fermions compactified on $S^3\times S^1$ in the large
$N$ limit.  The large $N$ limit is crucial to our analysis, because it
allows us to study phase transitions in finite spatial volume.  We took
advantage of the small $S^3$ limit  $R\Lambda \to 0$ to study the
partition function using perturbation theory, following
Refs.~\cite{Sundborg:1999ue,Polyakov:2001af,Aharony:2003sx,Hollowood:2009sy}.
The flow of singularities carries information about the appearance of Hagedorn
behavior and a confinement/deconfinement phase transition as the graded
partition function is deformed into the thermal partition functions of pure 
Yang-Mills theory. 

We observed two basic mechanisms for the transition. In the $n_f=1$ theory
(SUSY YM) singularities flow in from $q=1$, corresponding to ${\beta} =0$. For
$n_f> 1$ singularities appear as complex conjugate pairs which collide
on the real line and then turn into two real singularities that move apart
as $m$ is increased beyond the critical value. These two singularities have
opposite residues, so that the associated Hagedorn growth cancels at the
critical mass.  Note that in either case the critical value of $m_c R$
is finite, so that the mechanism which leads to intriguing Bose-Fermi
cancellation and absence of Hagedorn growth in $ \tilde\rho(E)$ persists
for $ mR < m_c R$. The fact that $m_c$ in $n_f =2$ theory is larger than
the one in the $n_f=1$ theory implies that Bose-Fermi cancellation is more
robust in the non-supersymmetric theory than the supersymmetric one under
mass deformation.

 An obvious question is what happens to the physics we have discussed as
$R\Lambda$ is increased from zero.  It would be interesting to study this
quantitatively in future work, but we can make some qualitative observations.
In this paper, we have interpreted the Hagedorn singularities as being
related to deconfinement.  This is only possible in the limit $R\Lambda
\to 0$, because (as we already mentioned in the introduction) the Hagedorn
temperature $\beta_H$ and the deconfinement temperature $\beta_d$ are only
equal in the zero-coupling limit $R\Lambda \to 0$.  For generic $R\Lambda$,
$\beta_d > \beta_H$, and the deconfinement phase transition is first-order.
Heuristically this is due to the fact that deconfinement is driven by a
change in the sign of the coefficient of $N^2$ in the free energy of the
quark-gluon plasma as a function of temperature, rather than by the dynamics
of the confined phase.  As a result, the interpretation of $\beta_H$ is that
it corresponds to the scale of the spinodal instability of the large $N$
confined phase:  the maximal temperature to which the confined phase can
be super-heated before local instabilities develop.  So at finite $R\Lambda$
a study of the large $N$ complex-temperature plane singularities of the
confined-phase partition function remains interesting, but its physical
interpretation changes.

A related question concerns the fate of the phase transitions as a function
of $m_c$ explored in this paper as $R\Lambda$ is increased from zero.  We
found that in general, $m_c \sim 1/R \sim \Lambda/(R\Lambda)$ when $R\Lambda
\to 0$. When $R\Lambda \to \infty$, it is known that the large $N$ critical
value of $m$ is actually $m/\Lambda=0$ in the infinite volume limit
\cite{Cherman:2018mya}.  As a result, it is tempting to guess that $m_c$
goes to zero monotonically with $1/(R\Lambda)$.  It would be nice to check
this hypothesis by an explicit calculation.

%%%%%%%%%%%%%%%%%%%%%%%%%%%%%%%%%%%%%%%%%%%%%%%%%%%%%%%%%%%%%%%%%%%%%%%%%
{\bf Acknowledgments.}  
We are grateful to O.~Costin for helpful discussions.  The work of S.~K.,
T.~S.~and M.~{\"U}.~is supported by the U.S. Department of Energy, Office
of Science, Office of Nuclear Physics under Award Number DE-FG02-03ER41260.

\vspace*{0.5cm}

\bibliography{small_circle}

\end{document}